\documentclass[12pt,a4paper]{article}
\usepackage{amsfonts,latexsym}
\usepackage{amsmath,amssymb}
\usepackage{graphicx,color}

\renewcommand{\thefootnote}{\fnsymbol{footnote}}

\begin{document}

\vspace{12mm}

\begin{center}
{{{\Large {\bf Radial perturbations of the  scalarized  black holes  \\ in Einstein-Maxwell-conformally coupled scalar theory }}}}\\[10mm]

{De-Cheng Zou$^{a,b}$\footnote{e-mail address: dczou@yzu.edu.cn} and Yun Soo Myung$^a$\footnote{e-mail address: ysmyung@inje.ac.kr}}\\[8mm]

{${}^a$Institute of Basic Sciences and Department  of Computer Simulation, Inje University, Gimhae 50834, Korea\\[0pt] }

{${}^b$Center for Gravitation and Cosmology and College of Physical Science and Technology, Yangzhou University, Yangzhou 225009, China\\[0pt]}
\end{center}
\vspace{2mm}

\begin{abstract}
We perform the stability analysis for the scalarized charged black holes obtained from Einstein-Maxwell-conformally coupled scalar (EMCS) theory by employing the radial perturbations.
The targeting black holes  include a single branch of  scalarized charged black hole with the coupling parameter $\alpha>0$  inspired by  the constant scalar hairy black hole
as well as infinite branches  of $n=0(\alpha \ge 8.019),1(\alpha \ge 40.84),2(\alpha \ge 99.89),\cdots$ scalarized charged black holes found  through the spontaneous scalarization on the Reissner-Nordstr\"om black hole. It turns out that  the black hole in the single branch and the $n = 0$ black
hole are  stable against radial  perturbations, while the $n = 1, 2$ excited black holes are
unstable in the EMCS theory with  both quadratic and exponential couplings.

\end{abstract}
\vspace{5mm}

\vspace{1.5cm}

\hspace{11.5cm}{Typeset Using \LaTeX}
\newpage
\renewcommand{\thefootnote}{\arabic{footnote}}
\setcounter{footnote}{0}


\section{Introduction}
No-hair theorem implies  that a black hole is completely described by  mass, electric charge, and angular momentum~\cite{Ruffini:1971bza}.
In this connection, we know well  that Maxwell and gravitational fields satisfy the Gauss-law outside the horizon.
It is interesting to note that a minimally coupled scalar  does not obey the Gauss-law and thus, a black hole
 could not  have a scalar hair in the Einstein-scalar theory~\cite{Herdeiro:2015waa}. On the other hand, introducing  the Einstein-conformally coupled scalar  theory leads to a secondary scalar hair around the BBMB (Bocharova-Bronnikov-Melnikov-Bekenstein) black hole~\cite{Bocharova:1970skc,Bekenstein:1974sf}.
This corresponds to  the first counterexample to the no-hair theorem for black holes.

Including  the Maxwell kinetic term with a scalar coupling  into  the Einstein-conformally coupled scalar  theory leads to the  EMCS theory.
The EMCS theory without scalar coupling   has admitted  the charged BBMB  black hole~\cite{Bekenstein:1974sf} and the constant scalar hairy  black hole~\cite{Astorino:2013sfa}.
It is emphasized   that  the former  implies  a secondary  scalar hair which  blows up on the horizon, while the latter has a constant scalar hair.
We would like to  mention that both of these  black holes are not considered as  a truly scalar hair, apart from the fact that  the former is unstable against the radial perturbation, while the latter is stable against the full perturbation.

On the other hand, scalarized black holes were found from the Einstein-Gauss-Bonnet-Scalar (EGBS) theory by introducing the quadratic and exponential couplings of a scalar
to the Gauss-Bonnet term like $\tilde{f}(\phi){\cal G}$~\cite{Doneva:2017bvd,Silva:2017uqg,Antoniou:2017acq}. In the EGBS theory including a non-minimally coupled scalar, the tachyonic instability of Schwarzschild black hole (a solution to the Einstein gravity) has triggered the spontaneous growth of a scalar (spontaneous scalarization) in the EGBS theory.
In this approach of spontaneous scalarization, a linearized scalar equation plays the crucial role to determine infinite branches of the $n=0,1,2,\cdots$ scalarized black holes.
In addition, scalarized charged black holes were obtained, through spontaneous scalarization~\cite{Herdeiro:2018wub},
from the instability of Reissner-Norstr\"{o}m (RN) black hole  in the Einstein-Maxwell-Scalar (EMS) theory~\cite{Myung:2018vug}.
Also, the spontaneous scalarization was discussed  in the EMS theory with general coupling~\cite{Fernandes:2019rez,Astefanesei:2019pfq,Blazquez-Salcedo:2020nhs}

Recently,  we  have obtained  a single branch of scalarized charged black holes  inspired by  the constant scalar hairy black hole as well as infinite branches  of $n=0(\alpha \ge 8.019),1(\alpha \ge 40.84),2(\alpha \ge 99.89),\cdots$ scalarized charged black holes with $\alpha$ coupling parameter found  through spontaneous scalarization in the EMCS theory~\cite{Zou:2019ays}.
These all are regarded really  as  charged black holes with scalar hair because they all  have  a primary scalar which takes  a finite value on the horizon.

Therefore, it is very important to investigate their stability  by considering  perturbations around scalarized charged black holes
in the EMCS theory with  both quadratic and exponential couplings.
In this work, we  would be better to  choose the radial perturbations  because the full perturbations around  scalarized charged black holes (numerical solutions) would  encounter some difficulty in achieving the  stability analysis for numerical black holes.

\section{EMCS theory}
The action for Einstein-Maxwell-conformally coupled scalar (EMCS) theory  takes the form
\begin{eqnarray}S_{\rm EMCS}=\frac{1}{16 \pi G}\int d^4 x\sqrt{-g}
\Big[R-\tilde{f}(\phi)F_{\mu\nu}F^{\mu\nu}-\beta\Big(\phi^2R+
6\partial_\mu\phi\partial^\mu\phi\Big)\Big],
\label{EMCS}
\end{eqnarray}
where $\tilde{f}(\phi)=1+\alpha \phi^2$ includes $\alpha \phi^2$ (quadratic  coupling with  coupling parameter $\alpha$) and  the last term corresponds to a conformally coupled scalar action with coupling parameter $\beta$. In section 5, we will introduce the exponential coupling of $\tilde{f}(\phi)=e^{\alpha \phi^2}$ as a nonlinear coupling to perform
the stability analysis of scalarized charged black holes.
In this work,  we choose $\beta=1/3$ and $G=1$ for simplicity because any $\beta>0$ would  not change the main results.
In the decoupling limit of $\alpha\to 0$, the above action reduces to the $\alpha=0$ EMCS theory which allowed  the constant scalar hairy  black hole and  charged BBMB  black hole.
The  Einstein equation  is derived from (\ref{EMCS}) as
\begin{equation} \label{nequa1}
G_{\mu\nu}=2(1+\alpha \phi^2) T^{\rm M}_{\mu\nu}+T^{\rm \phi}_{\mu\nu},
\end{equation}
where the energy-momentum tensors for Maxwell theory and  conformally coupled scalar theory  are  given, respectively,  by
\begin{eqnarray} \label{equa2}
T^{\rm M}_{\mu\nu}&=&F_{\mu\rho}F_{\nu}~^\rho- \frac{F^2}{4}g_{\mu\nu},\label{trace} \\
T^{\rm \phi}_{\mu\nu}&=&\beta\Big[\phi^2G_{\mu\nu}+g_{\mu\nu}\nabla^2(\phi^2)-\nabla_\mu\nabla_\nu(\phi^2)+6\nabla_\mu\phi\nabla_\nu\phi-3(\nabla\phi)^2g_{\mu\nu}\Big]\nonumber
\end{eqnarray}
with $F^2=F_{\rho\sigma}F^{\rho\sigma}$ and $\beta= \frac{1}{3}$.
Here, we observe the traceless condition of $T^{{\rm M} \mu}_\mu=0$ for Maxwell field. The Maxwell equation is given by
\begin{equation} \label{maxwell-eq}
\nabla^\mu F_{\mu\nu}=-2\alpha \phi \nabla_{\nu}(\phi) F^2.
\end{equation}
On the other hand, the scalar
equation is given by
\begin{equation} \label{ascalar-eq}
\nabla^2\phi-\frac{1}{6}R\phi-\frac{\alpha}{6\beta}  F^2 \phi=0.
\end{equation}
Considering  the trace of the Einstein equation (\ref{nequa1}) together with (\ref{ascalar-eq}) implies  a non-vanishing Ricci scalar given by
\begin{equation} \nonumber
R=-\alpha \phi^2  F^2.
\end{equation}
Finally, we obtain  a non-minimally coupled scalar equation
\begin{equation} \label{scalar-eq}
\nabla^2\phi+\frac{\alpha}{6}\Big[\phi^2-\frac{1}{\beta}\Big]  F^2 \phi=0,\quad \beta=\frac{1}{3}.
\end{equation}
In case of $\alpha=0$, one finds  a  minimally coupled scalar equation ($\nabla^2\phi=0$) with $R=0$ which admitted the  charged BBMB  black hole~\cite{Bekenstein:1974sf} and the constant scalar hairy  black hole~\cite{Astorino:2013sfa}. However, it is important to note that  the case of $\phi=\pm\sqrt{3}$ and $\alpha\not=0$  is  not suitable for admitting any charged black hole because its Einstein equation is ill-defined as
 \begin{equation}
 (1-\phi^2/3)G_{\mu\nu}=2(1+\alpha \phi^2)T^M_{\mu\nu}\to 0=2(1+\alpha \phi^2)T^M_{\mu\nu}. \label{WE-eq}
\end{equation}

\section{Scalarized charged black holes}
Before we proceed, we would like to introduce an analytical solution of the RN  black hole without scalar hair found in the EMCS theory
\begin{eqnarray}
&&ds^2_{\rm RN}=-f(r) dt^2+\frac{dr^2}{f(r)}+r^2d\Omega^2_2, \nonumber \\
&&f(r)=1-\frac{2M}{r}+\frac{Q^2}{r^2},~~\bar{\phi}=0,~~\bar{A}_t=\frac{Q}{r}-\frac{Q}{r_+}. \label{RN}
\end{eqnarray}
We briefly mention the instability issue of the RN black hole in the EMCS theory because it is the starting point of spontaneous scalarization.
The linearized  equation for  the perturbed scalar $\delta \phi$ is given by
\begin{equation} \label{linRNs}
\bar{\nabla}^2\delta \phi+\frac{\alpha \bar{F}^2}{2} \delta \phi=0,
\end{equation}
which determines  the instability of the RN black hole. At this stage, it is noted that $\beta=1/3$ has nothing to do with the instability of the RN black hole.
The last term in (\ref{linRNs}) is  an  effective mass term which plays a role of the tachyonic mass.
Thus, this term develops instability of the RN black hole, depending on the coupling parameter $\alpha$.
We do not wish to  display  the stability analysis of the RN black hole in the EMCS theory explicitly
 because it is  exactly the same  for the EMS theory~\cite{Myung:2018vug}.
We have obtained  the infinite branches of solutions labeled by $n=0(\alpha\ge 8.019),~n=1(\alpha\ge 40.84),~n=2(\alpha \ge 99.89),\cdots$ scalarized charged black holes
with $q=Q/M=0.7$~\cite{Zou:2019ays}, through the spontaneous scalarization,  from the static version of (\ref{linRNs}).
Explicitly, these were determined by the equation~\cite{Herdeiro:2018wub}
\begin{equation} \label{anal-br}
{}_2F_1\Big[0.5(1 - \sqrt{1 - 4 \alpha}),0.5(1 + \sqrt{1 - 4 \alpha}), 1, \frac{q^2}{2(q^2-1-\sqrt{1-q^2})}\Big]|_{q=0.7}=0
\end{equation}
with ${}_2F_1[\cdots]$ the hypergeometric function.

Also, it is interesting to introduce  the constant scalar hairy black hole obtained from the $\alpha=0$ EMCS theory without scalar coupling~\cite{Astorino:2013sfa,Khodadi:2020jij}
\begin{eqnarray}
&&ds^2_{\rm csbh}=-\tilde{f}(r)dt^2+\frac{dr^2}{\tilde{f}(r)}+r^2d\Omega^2_2, \nonumber \\
&&\tilde{f}(r)=1-\frac{2m}{r}+\frac{Q^2+q^2_s}{r^2},~~\bar{\phi}_c=\sqrt{\frac{3q^2_s}{q^2_s+Q^2}},~~\bar{A}_t=\frac{Q}{r}-\frac{Q}{r_+}, \label{cshbh}
\end{eqnarray}
where $q_s$ does not represent a truly scalar charge $Q_s$ existing in the $\alpha\not=0$ EMCS theory.
This solution could be obtained by solving the Einstein equation of $(1-\phi^2/3)G_{\mu\nu}=2T^M_{\mu\nu}$ only  for $\phi\not=\pm\sqrt{3}$ from the $\alpha=0$ EMCS theory.
If $\phi=\pm\sqrt{3}$, one finds a wrong Einstein equation (\ref{WE-eq}). Furthermore, imposing $\bar{\phi}_c=0(q_s=0)$ on (\ref{cshbh}) reduces to the RN black hole in (\ref{RN}).

It seems that (\ref{cshbh}) is not relevant to the spontaneous scalarization  because there is no effective mass term ($\bar{\nabla}^2\delta \phi=0$) and thus, it would not suffer from a tachyonic instability. Actually, it was known that  this black hole was stable against full perturbations in the $\alpha=0$ EMCS theory~\cite{Zou:2019ays}, suggesting a single branch without other branches.
If the constant scalar hairy solution (\ref{cshbh}) is really found from  the $\alpha\not=0$ EMCS theory, one has to consider the linearized scalar equation (instead of $\bar{\nabla}^2\delta \phi=0$)
\begin{equation} \label{lincshbh}
\bar{\nabla}^2\delta \phi+\frac{\alpha \bar{F}^2}{2}\Big[\bar{\phi}_c^2-1\Big]\delta \phi+\frac{\alpha(\bar{\phi}_c^2-3)\bar{\phi}_c}{6}\delta F^2=0,
\end{equation}
where
\begin{equation}
\delta F^2=-2\Big(\bar{F}_{\rho\sigma}f^{\rho\sigma}-\bar{F}_{\kappa\rho}\bar{F}^{\kappa}~_{\sigma} h^{\rho\sigma}\Big) \nonumber
\end{equation}
with the perturbed Maxwell field $f_{\mu\nu}=\partial_\mu a_\nu-\partial_\nu a_\mu$ and the perturbed metric tensor $h_{\mu\nu}$.
Eq.(\ref{lincshbh}) may provide the instability of the constant scalar hairy black hole. However, it is very important to note that the constant scalar hairy black hole (\ref{cshbh}) is not a solution to the $\alpha\not=0$ EMCS theory, but a solution to the $\alpha=0$ EMCS theory without scalar coupling. Hence, the usage of (\ref{lincshbh}) might mislead to analyzing the instability of the constant scalar hairy black hole, which may allow other branches as in (\ref{anal-br}).

Anyway, introducing the solution (\ref{cshbh})  is meaningful because it plays a role of the guideline for constructing the single branch of scalarized charged black holes in  the $\alpha\not=0$ EMCS theory.
This single branch is never found from  the EMS theory without conformally coupled scalar term.  In addition, we would like to mention that
the $\alpha=0$ EMCS theory is not invariant under conformal transformation because of the presence of the Einstein-Hilbert term [the first term in (\ref{EMCS})].
In this theory, one obtains  $R=0$ and $\nabla^2 \phi=0$ from breaking of the conformal symmetry.

Let us  assume the metric and fields to find scalarized charged black holes
\begin{eqnarray}
&&ds^2_{\rm scbh}=-N(r)e^{-2\delta(r)}dt^2+\frac{dr^2}{N(r)}+r^2d\Omega^2_2, \nonumber \\
&&N(r)=1-\frac{2m(r)}{r},~~\bar{\phi}=\phi(r),~~\bar{A}_t=v(r). \label{sRN}
\end{eqnarray}
Substituting (\ref{sRN}) into Eqs.(\ref{nequa1}), (\ref{maxwell-eq}), and (\ref{scalar-eq}), one finds four equations for $m(r),\delta(r),v(r)$, and $\phi(r)$ as
\begin{eqnarray}
&&3e^{2\delta}r^2\alpha\phi(\phi^2-3)v'^2-18(m-m'r)\phi
-r(r-2m)(9+\phi^2)\phi''\nonumber\\
&&-(r-2m)\Big[\phi(\phi^2-3)\delta'
+\Big(18+r(\phi^2-9)\delta'\Big)\phi'-2r\phi\phi'^2\Big]=0,\label{neom1}\\
&&3e^{2\delta}r^2(1+\alpha\phi^2)v'^2+2(r-2m)(\phi^2-3)\delta'
+2\phi\Big(3m-2r+r(r-2m)\delta'\Big)\phi'\nonumber\\
&&-3r(r-2m)\phi'^2+2m'(\phi^2+r\phi\phi'-3)=0,\label{neom2}\\
&&\Big(2+r\delta'+\frac{2r\alpha\phi\phi'}{1+\alpha\phi^2}\Big)v'+rv''=0,\label{neom3}\\
&&r\phi\phi''-2r\phi'^2+\delta'(\phi^2-3+r\phi\phi')=0, \label{neom4}
\end{eqnarray}
where the prime ($'$) denotes differentiation with respect to $r$.

The constant scalar hairy solution (\ref{cshbh}) is different from the RN solution (\ref{RN}) in the sense that the former has a constant scalar hair, while
the latter has no such a scalar hair. These will be   two bases for two tracks.
We have two tracks to obtain scalarized charged black holes. One is to derive the scalarized charged black holes  by considering the  constant scalar hairy black hole (\ref{cshbh})
because the EMCS theory contains the conformally coupled scalar term. In this case, we could not use (\ref{lincshbh}) to perform the instability analysis of the  constant scalar hairy black hole   because it  is
not a solution to the $\alpha\not=0$ EMCS theory. If this is the case, one may use (\ref{lincshbh}) to generate infinite branches of scalarized charged black holes, depending on the coupling parameter $\alpha$. Here, one may use a minimally coupled linearized equation $\bar{\nabla}^2 \delta \phi=0$, which is stable and provides a single branch of scalarized charged black holes in the $\alpha\not=0$ EMCS theory.
 The other is a conventional approach to deriving infinite branches of scalarized charged black holes from
the onset of spontaneous scalarization based on the instability of the RN black hole.

\subsection{Scalarized charged black holes in the single  branch }
In this section, we briefly review how to derive the scalarized charged black holes inspired by  the  constant scalar hairy black hole (\ref{cshbh}).
Implementing  an outer horizon located at $r=r_+$,  one may  introduce  an
approximate solution to (\ref{neom1})-(\ref{neom4}) in the near-horizon region
\begin{eqnarray}\label{nexpr}
&&m(r)=\frac{r_+}{2}+m_1(r-r_+)+\cdots,\label{aps-1}\\
&&\delta(r)=\delta_0+\delta_1(r-r_+)+\cdots,\label{aps-2}\\
&&\phi(r)=\phi_0+\phi_1(r-r_+)+\cdots,\label{aps-3}\\
&&v(r)=v_1(r-r_+)+\ldots,\label{aps-4}
\end{eqnarray}
where the coefficients are finite and determined  by
\begin{eqnarray}
&&m_1=\frac{[(\alpha\phi_0^2(\phi_0^2-12)-9]Q^2}{6r_+^2(\phi_0^2-3)(1+\alpha\phi_0^2)^2},\nonumber\\
&&\delta_1=\frac{\alpha\phi_0^2 Q^2(\phi_0^2-3)}{2r_+(1+\alpha\phi_0^2)\Big[Q^2(9-\alpha\phi_0^2(\phi_0^2-12))
+3r_+^2(\phi_0^2-3)(1+\alpha\phi_0^2)^2\Big]^2} \label{nceof} \\
&&\times \Big[12r_+^2(\phi_0^2-3)(1+\alpha\phi_0^2)^3+
Q^2\{18+\alpha(27+(48+63\alpha)\phi_0^2+(6\alpha-7)\phi_0^4
-3\alpha\phi_0^6)\}\Big],\nonumber \\
&&\phi_1=\frac{\alpha\phi_0 Q^2(\phi_0^2-3)^2}{r_+Q^2\left(9-\alpha\phi_0^2(\phi_0^2-12)\right)
+3r_+^3(\phi_0^2-3)(1+\alpha\phi_0^2)^2}, \nonumber \\
&&v_1=-\frac{e^{-\delta_0}Q}{r_+^2(1+\alpha\phi_0^2)}.\nonumber
\end{eqnarray}
It is interesting to note that $\delta_1$ takes a complicated form because of a conformal coupling term $\phi^2 R$.

Before we proceed, we mention that (\ref{nceof}) represents the most general behavior of the black hole in the near-horizon region.
It may include RN black hole, constant scalar hairy black hole, and scalarized charged black hole with scalar hair.

Firstly, let us consider  what happens for $\phi_0=\pm\sqrt{3}$.
Here, one finds an unwanted case  that $m_1\to \infty,~\delta_1\to 0,~\phi_1\to 0$, where $m_1$ blows up.

In  case of $\alpha=0$ and $\phi_0\not=\pm\sqrt{3}(Q\not=0)$, the above coefficients reduce to those for the constant scalar hairy black hole exactly
\begin{eqnarray}
&& m_1=-\frac{9Q^2}{6r_+^2(\phi_0^2-3)}=\frac{Q^2+q_s^2}{2r_+^2},~~\delta_0=\delta_1=0, \nonumber \\
&&\phi_0=\bar{\phi}_c=\sqrt{\frac{3q^2_s}{Q^2+q^2_s}},~~\phi_1=0,~~v_1=-\frac{Q}{r_+^2}, \label{cshF}
\end{eqnarray}
where the case of $\phi_0=0(q_s=0)$ leads to the RN black hole.

The near-horizon solution (\ref{nceof}) involves  two essential  parameters of  $\phi_0=\phi(r_+,\alpha)$ and $\delta_0=\delta(r_+,\alpha)$ on the horizon, which can be
determined  by  matching  (\ref{aps-1})-(\ref{aps-4}) with an asymptotic  solution in the far-region as
\begin{eqnarray}\label{insolC}
m(r)&=&M-\frac{3Q_s^2(1+\phi_\infty^2)}{2(\phi_\infty^2-3)^2r}
-\frac{Q^2(2\alpha\phi_\infty^4-15\alpha\phi_\infty^2-9)}{6(\phi_\infty^2-3)(1+\alpha\phi_\infty^2)^2r}
-\frac{MQ_s\phi_\infty}{(\phi_\infty^2-3)r}\cdots, \nonumber \\
\delta(r)&=&\frac{2Q_s\phi_\infty}{(\phi_\infty^2-3)r}+\cdots, \label{shar1} \\
 \phi(r)&=&\phi_\infty+\frac{Q_s}{r}+\cdots, \nonumber \\
 v(r)&=&\Psi+\frac{Q}{(1+\alpha\phi_\infty^2)r}+\cdots, \nonumber
\end{eqnarray}
where $M$, $\Psi$, and $Q_s$ are the ADM mass, the electrostatic potential at infinity, and the scalar charge.
Here, we note that $m(r)$ and $\delta(r)$ blow up for $\phi_\infty=\pm\sqrt{3}$.

In case of $\alpha=0$ and $Q_s=0$, the above reduces to those  for the constant scalar hairy black hole exactly
\begin{eqnarray} \label{cshar}
m(r)=M+\frac{9Q^2}{6(\phi^2_\infty-3)r}=M-\frac{Q^2+q^2_s}{2r},~\delta(r)=0,~\phi(r)=\phi_\infty=\bar{\phi}_c,~v(r)=\Psi+\frac{Q}{r},
\end{eqnarray}
which plays a role of the guideline for constructing scalarized charged black holes in the single branch.
From (Left) Fig. 1, we obtain  a scalar hair $\phi_0(\alpha)=\bar{\phi}(r_+,\alpha)$ on the horizon in the single branch of scalarized charged black holes with a positive $\alpha$, implying no other branch of scalarized charged black holes.
\begin{figure*}[t!]
   \centering
   \includegraphics{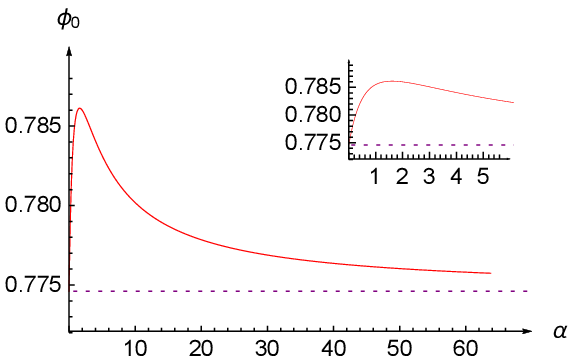}
\hfill%
  \includegraphics{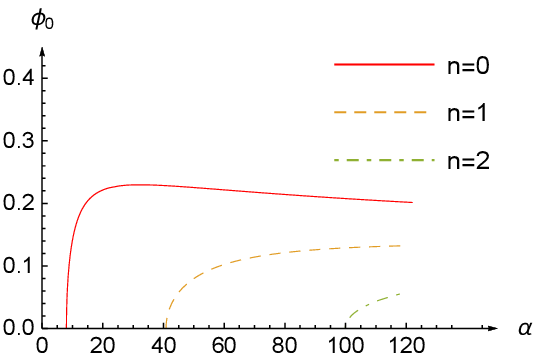}
\caption{(Left) Plot of  scalar hair $\phi_0(\alpha)=\bar{\phi}(r_+,\alpha)\ge \bar{\phi}_c$ on the horizon as function of the coupling parameter $\alpha\in[0,\infty)$, showing a single branch.
The dashed line denotes the constant scalar hairy black hole with $\phi_0=\bar{\phi}_c=0.7746=\phi_\infty$.
 (Right) Plots of  $\phi_0(\alpha)=\bar{\phi}(r_+,\alpha)$  for the first three branches among infinite branches.
  The $n=0$ branch starts from the first bifurcation point at $\alpha=8.019$, and  $n=1$ and 2 branches start from the second point at $\alpha=40.84$ and from the third point at $\alpha=99.89$. }
\end{figure*}

Explicitly, we wish to show a (numerical) scalarized charged black hole with $\alpha=63.75$  in Fig.  2.
$N(r)$ and $\tilde{f}(r)$ represent metric function for scalarized charged black hole and constant scalar hairy black hole, respectively.
The magnification in the left picture indicates  an enlarged decrease  of scalar hair $\phi(r)\ge \bar{\phi}_c$, showing a clear difference  from a constant hair $\bar{\phi}_c=0.7746$ for the constant scalar hairy black hole~\cite{Zou:2019ays}.  It is worth noting  that  this scalar hair is not constant and does not blow up on the horizon and thus, it is surely a primary one.
\begin{figure*}[t!]
   \centering
   \includegraphics{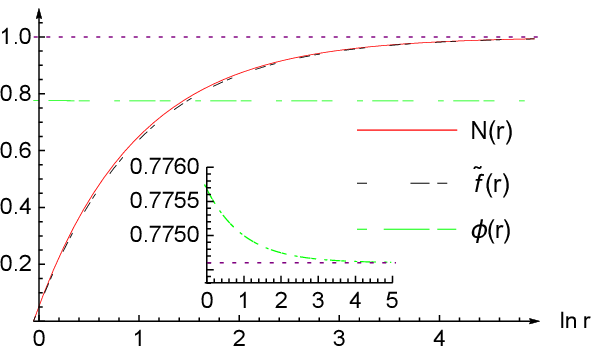}
         \hfill%
    \includegraphics{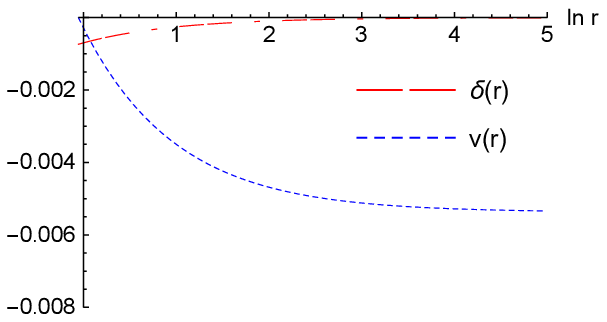}
\caption{ (Left) Plot of a scalarized black hole with  $\alpha=63.75$, compared to the constant scalar hairy black hole [$\tilde{f}(r),~\bar{\phi}_c=0.7746$]. The horizon is located at $\ln r=\ln r_+=-0.0542$. The right picture indicates that  $\delta(r)$ is negative [$\delta(r)=0$ for the constant scalar hairy black hole]
and $v(r)$ is a negative function.}
\end{figure*}

\subsection{$n=0,1,2,\cdots$ scalarized charged black holes}

In this case, an  approach to finding infinite black hole solutions through the spontaneous scalarization is the nearly same in the previous one except that  the
the asymptotic solution in the far-region is given by
\begin{eqnarray}\label{insol}
m(r)&=&M-\frac{3Q^2+Q_s^2}{6r}+\ldots, \quad \phi(r)=\frac{Q_s}{r}+\cdots, \nonumber \\
\delta(r)&=&\frac{Q_s^2[2Q_s^2-6M^2+3Q^2(2+\alpha)]}{108r^4}+\cdots,\quad
v(r)=\Psi+\frac{Q}{r}+\cdots,
\end{eqnarray}
whose limit of $\phi=0(Q_s=0)$ corresponds to  the RN black hole.
Actually, (\ref{insol}) could be obtained when imposing $\phi_\infty=0$ in (\ref{shar1}). Thus, any asymptotic scalar hairs are absent in the far-region, differing  from  $\phi_\infty=\bar{\phi}_c$ found in the single branch of scalarized charged black holes. Here, we find that the blow-up points for $m(r)$ and $\delta(r)$ in (\ref{shar1}) disappear.
From (Right) Fig. 1, we have obtained  the three branches of solutions labeled by $n=0(\alpha\ge 8.019),~n=1(\alpha\ge 40.84),~n=2(\alpha \ge 99.89)$ scalarized charged black holes. At this stage, we wish to note that the appearance of these black holes with scalar hair is closely connected to the onset of scalarization based on the instability of RN black holes determined by the linearized scalar equation (\ref{linRNs}) with $q=0.7$.
Explicitly, we choose the  horizon radius $r_+=0.857$ and electric charge $Q=0.35$
to  construct the $n=0$ scalarized charged black hole with $\alpha=65.25$ shown in Fig. 3. Here $\phi(r)$ starts with $\phi_0(r_+=0.857,\alpha=65.25)=0.2196$ and its asymptotic value is  zero.
\begin{figure*}[t!]
   \centering
   \includegraphics{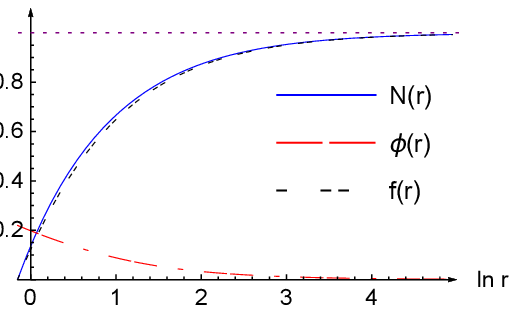}
      \hfill%
      \includegraphics{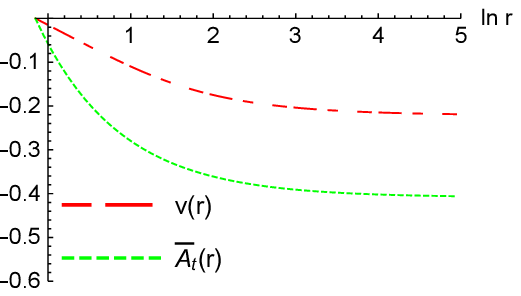}
        \hfill%
    \includegraphics{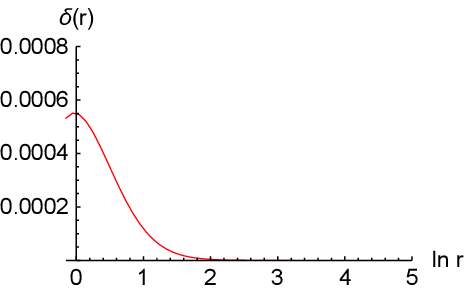}
\caption{Graphs of a scalarized charged black hole with $\alpha=65.25$  in the  $n=0$ branch. Here $f(r)$ and  $\bar{A}_t$  represent the metric function  and vector potential  for the RN black hole with $\delta(r)=0$.
 We plot all figures  in terms of  `$\ln r$' and thus,
 the horizon is located at $\ln r=\ln r_+=-0.154$.}
\end{figure*}

\section{Stability of scalarized charged black holes}
Before we proceed, we have to mention that it is not an easy task to carry out the stability of scalarized charged black holes because these black holes
come out as not analytic solutions but numerical solutions. In order to develop the stability analysis, one needs to obtain hundreds of numerical solutions depending on the coupling constant $\alpha$  in the each branch.
Also, the full (axial+polar) perturbations require  a complicated decoupling process because the linearized  EMCS theory contains five physically propagating modes on these black hole background. In addition, we note that  the $l=0$ ($s$-mode) scalar propagation determines mainly the stability of these black holes. In the conformal coupling theory
(EMCS theory), it is would be better to choose the radial (spherically symmetric) perturbations starting with two metric and one vector perturbations  which are  regarded as a simpler version of the polar perturbation as far as the scalar perturbation is concerned.

Let us  introduce  the radial  perturbations around the scalarized black holes as
\begin{eqnarray}
&&ds^2_{\rm rad-p}=-N(r)e^{-2\delta(r)}(1+\epsilon H_0)dt^2+\frac{dr^2}{N(r)(1+\epsilon H_1)}
+r^2\left(d\theta^2+\sin^2\theta d\varphi^2\right),\nonumber\\
&& \phi(t,r)=\phi(r)+\epsilon \frac{\Phi(t,r)}{r}, \quad
F_{rt}(t,r)=v'(r)+\epsilon\delta v(t,r),\label{paction}
\end{eqnarray}
where  $N(r)$, $\delta(r)$, $\phi(r)$, and $v(r)$ represent a
scalarized charged  black hole background, while $H_0(t,r)$, $H_1(t,r)$, $\Phi(t,r)$, and  $\delta v(t,r)$
denote four perturbed fields around the scalarized black hole background. From now on, we confine ourselves to analyzing the $l=0$($s$-mode)  propagation, implying that higher angular momentum modes ($l\not=0$) are excluded.
In this case, all perturbed fields except the perturbed  scalar  $\Phi$ may belong to redundant fields.

According to Appendix, we find
the decoupled scalar equation for testing  the stability of scalarized charged black holes as
\begin{eqnarray}
\Big[g^2(r)\frac{\partial^2\Phi}{\partial t^2}\Big]-\frac{\partial^2\Phi}{\partial r^2}
+C_1(r)\frac{\partial\Phi}{\partial r}+U(r)\Phi=0,\quad g(r)=\frac{e^{\delta(r)}}{N(r)}, \label{nrad-eq}
\end{eqnarray}
where  $C_1(r)$ and $U(r)$ are expressed in Appendix.

Introducing  a further separation of  $\Phi(t,r)=\delta\phi(r)e^{-i\omega t}$,
we obtain the Schr\"{o}dinger-like equation from (\ref{nrad-eq})
\begin{equation}
\frac{d^2Z(r)}{dr_*^2}+[\omega^2-V(r,\alpha)]Z(r)=0, \label{perteq}
\end{equation}
where $r_*$ is a tortoise coordinate to extend from $r\in[r_+,\infty]$ to $r_*\in[-\infty,\infty]$ and  $Z(r)$ is  a redefined scalar,
expressed by
\begin{eqnarray}
r_*=\int^\infty_{r_+}g(r)dr,~~Z(r)=\frac{\delta\phi(r)}{C_0(r)}.
\end{eqnarray}
Here the potential takes the form
\begin{eqnarray}
V(r,\alpha)=\frac{U(r)-C'_1(r)}{g^2(r)}+\frac{C_1g'(r)+g''(r)}{g^3(r)}-\frac{[2g'(r)]^2}{g^4(r)}.
\end{eqnarray}
We  point out that $C_0(r)$ is the solution to
the  differential equation
\begin{eqnarray}
[\ln C_0(r)]'=C_1(r)-[\ln g(r)]'
\end{eqnarray}
which means that it is difficult to solve for $C_0(r)$ analytically because of a complicated form $C_1(r)$.
\begin{figure*}[t!]
   \centering
   \includegraphics{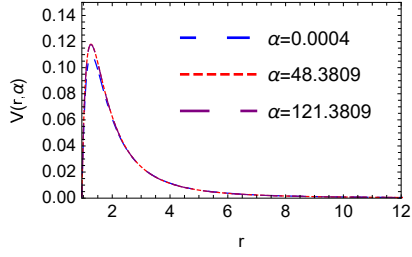}
\caption{Scalar potential $V(r,\alpha)$  around scalarized charged black holes in the single branch.
All  potentials are positive definite outside the horizon for $\alpha>0$, implying the stable black hole.    }
\end{figure*}
\begin{figure*}[t!]
   \centering
   \includegraphics{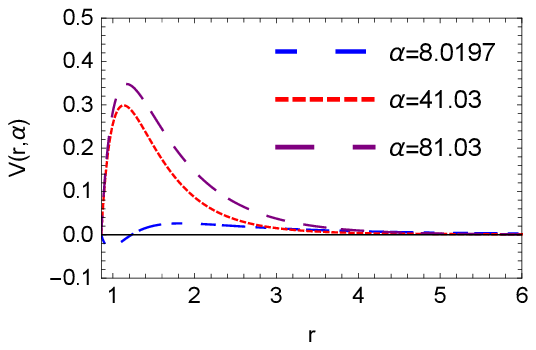}
    \hfill%
    \includegraphics{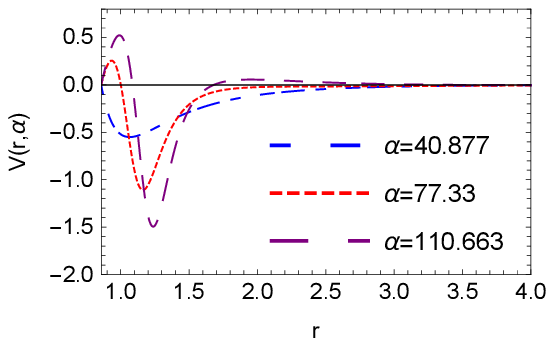}
    \hfill%
    \includegraphics{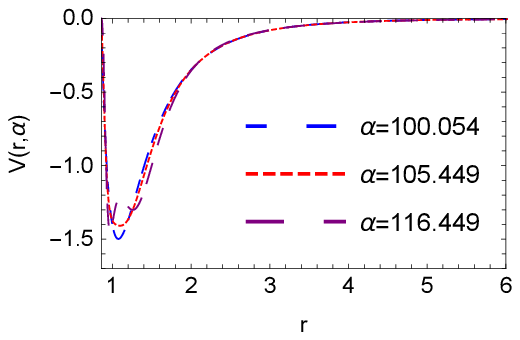}
\caption{Scalar potentials $V(r,\alpha)$  around $n=0$ (Left: $\alpha\ge8.019$), 1 (Middle: $\alpha\ge40.84$), 2 (Right: $\alpha\ge99.89$) black holes in the infinite branches.  }
\end{figure*}
Usually, a positive definite potential $V(r,\alpha)$ excluding any negative region guarantees the stability of the black hole.
It is clear from Fig. 4 that the potential $V(r,\alpha)$ around the black hole in the single branch  are positive definite outside the horizon for $\alpha>0$, indicating  the stability.
In this case, we do not need to perform a further analysis for stability.
We inform from (Left) Fig. 5  that the potential around the $n=0$ black hole indicates  large positive region outside the horizon, suggesting  the stability.
On the other hand, from (Middle, Right) Fig. 5,   the potentials around the $n=1,2$ black holes show large negative regions outside  the  horizon, showing the instability.
In addition, a sufficient condition for instability is given by $\int^\infty_{r_+} dr[g(r)V]<0$ in accordance with the existence of unstable modes~\cite{Dotti:2004sh}.
However, a potential with negative region near the horizon  whose integral ($\int^\infty_{r_+} dr[g(r)V]>0$) is positive might not exclude the existence of stable modes.

Importantly, to determine the (in)stability of the black hole, we have to solve (\ref{perteq}) numerically by imposing an appropriate boundary condition  that $Z(r)$ has an outgoing wave at infinity and an ingoing wave on the horizon:
$Z(r)\sim e^{i\omega r_*}$ at $r_*\to\infty$ and $ Z(r)\sim e^{-i\omega r_*}$ at $r_*\to-\infty$.
If one finds an exponentially growing mode of $e^{\Omega t}(\omega=i \Omega)$, the corresponding black hole is unstable against the  perturbation.
The linearized  scalar equation (\ref{perteq}) around the scalarized charged black hole in  the $n=0,1,2$ scalarized charged  black holes may allow   either a stable (decaying) mode with $\Omega<0$ or an unstable (growing) mode with $\Omega>0$.
In case of unstable modes, we may solve (\ref{perteq}) numerically  with a boundary condition that
$Z(r)= 0$ at $r_*=\infty$ and $ Z(r)= 0$ at $r_*=-\infty$. We find from Fig. 6 that  the $n=0$ black hole is stable against the $l=0$-scalar mode,
whereas    the $n=1, 2$ black holes are  unstable against the $l=0$-scalar mode.
\begin{figure*}[t!]
   \centering
   \includegraphics{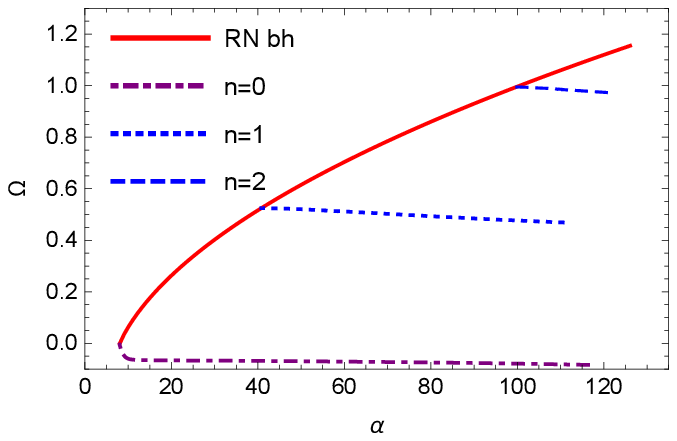}
    \hfill%
    \includegraphics{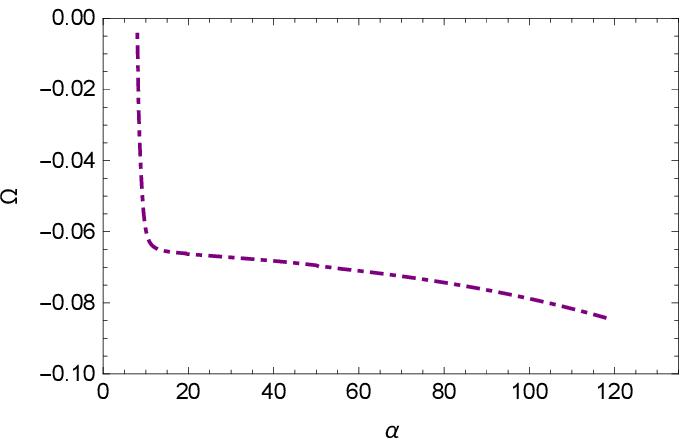}
\caption{(Left) Plots of $\Omega$ as functions of  $\alpha$ for $l=0$-scalar mode around the $n=0(\alpha\ge8.019)~,1(\alpha\ge40.84),~2(\alpha\ge99.89)$ black holes.
The positive $\Omega$ for $n=1,2$ imply  unstable black holes, while the negative $\Omega$ for $n=0$ shows a stable  black hole.
 A red curve with $q=0.7$ denotes the positive $\Omega$ around the unstable RN black hole for $\alpha>8.019$. (Right) The enlarged picture shows the  negative $\Omega$  for $\alpha\ge 8.019$ in the $n=0$ black hole clearly. }
\end{figure*}

\section{Stability of scalarized charged black holes with exponential coupling}
The tachyonic instability of black holes without scalar hair triggering the spontaneous scalarization would be quenched by introducing a nonlinear scalar coupling term.
In case of  the EGBS theory, the scalarized black hole solutions  in the pure quadratic coupling are always unstable~\cite{Silva:2017uqg,Blazquez-Salcedo:2018jnn},
whereas  the scalarized black hole solutions in the nonlinear coupling models could be stable~\cite{Blazquez-Salcedo:2018jnn,Minamitsuji:2018xde,Silva:2018qhn} because the nonlinear scalar terms may become more important as time passes and quench a growth of the scalar field.
Furthermore,  it was shown that  the stability analysis of scalarized charged black holes makes no difference  between   quadratic coupling for the EMS theory~\cite{Myung:2019oua} and  exponential coupling~\cite{Myung:2018jvi}.   This means that the $n=0$ black hole is stable, while the $n=1,2$ excited black holes are unstable, irrespective of coupling.

In this section, we wish to  analyze the stability of scalarized charged black holes obtained from  the EMCS theory with exponential coupling $\tilde{f}(\phi)=e^{\alpha \phi^2}$
whose limit of small amplitude reduces to the action (\ref{EMCS})  of  the EMCS theory. We  expect to derive a stable scalarized charged black hole.
Here,  we briefly describe  scalarized charged black holes and their stability analysis without mentioning  mathematical expressions
because their mathematical expressions are similar to those in the EMCS theory (with quadratic coupling).
\begin{figure*}[t!]
   \centering
   \includegraphics{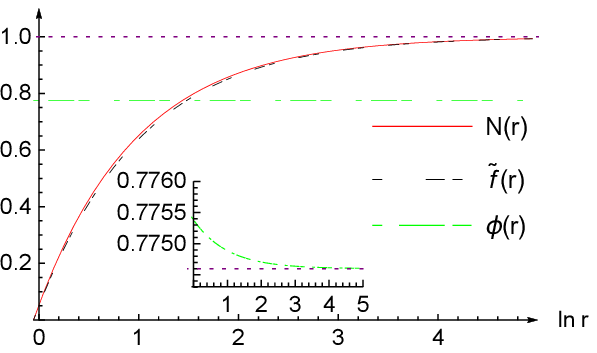}
         \hfill%
    \includegraphics{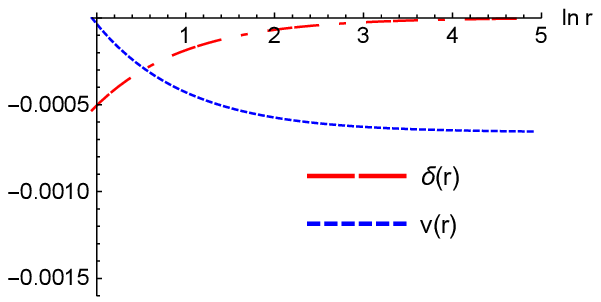}
\caption{ (Left) Plot of a scalarized black hole with  $\alpha=9.6069$ in the single branch for exponential coupling. The horizon is located at $\ln r=\ln r_+=-0.0542$.
$\tilde{f}(r)$ and $\bar{\phi}_c=0.7746 $ represent the constant scalar hairy black hole.   The right picture indicates  $\delta(r)$ being a negative function and $v$ being a negative function .}
\end{figure*}
First of all, we obtain a scalarized charged black hole solution in the single branch (see Fig. 7) and a scalarized charged black solution in the $n=0$ branch (see Fig. 8)
by following Sec. 3 after replacing $1+\alpha\phi^2$ by exponential coupling $e^{\alpha \phi^2}$.
In order to carry out their stability analyses, we redo the analysis in Sec. 4 after replacing  after replacing $1+\alpha\phi^2$ by $e^{\alpha \phi^2}$.

Now, we obtain potential $V(r,\alpha)$ [Fig. 9] in the single branch and $V(r,\alpha)$ [Fig. 10] in the $n=0,~1,~2$ branches, which are very similar to the previous potentials in Figs. 4 and 5. The single branch and $n=0$ branch correspond to potentials with positive region, implying the stability.
The other cases of $n=1$ and $2$ suggest instability of scalarized charged black holes   because  negative regions are larger than positive regions, leading to the sufficient condition for instability given by $\int^\infty_{r_+} dr[g(r)V]<0$.

\begin{figure*}[t!]
   \centering
   \includegraphics{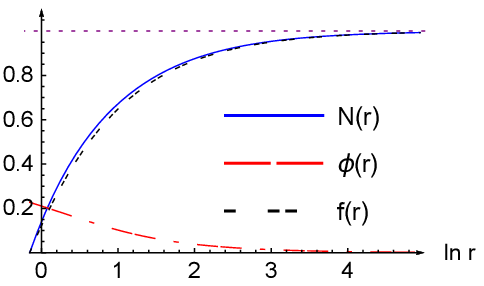}
      \hfill%
      \includegraphics{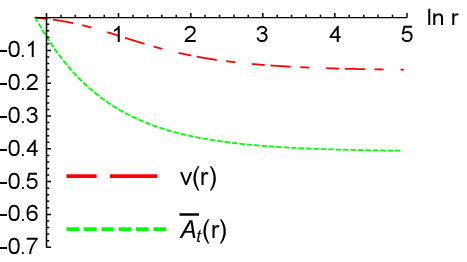}
        \hfill%
    \includegraphics{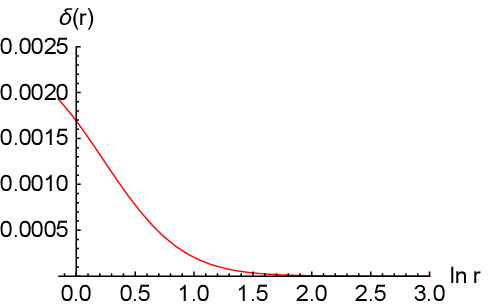}
\caption{Graphs of a scalarized charged black hole with $\alpha=58.4167$  in the  $n=0$ branch for exponential coupling. Here $f(r)$ and  $\bar{A}_t$  represent the metric function  and vector potential  for the RN black hole with $\delta_{\rm RN}(r)=0$.
 The horizon is located at $\ln r=\ln r_+=-0.154$.}
\end{figure*}
\begin{figure*}[t!]
   \centering
   \includegraphics{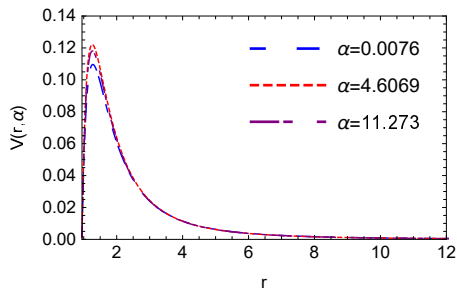}
\caption{Scalar potential $V(r,\alpha)$  around scalarized charged black holes in the single branch for exponential coupling.   }
\end{figure*}
\begin{figure*}[t!]
   \centering
   \includegraphics{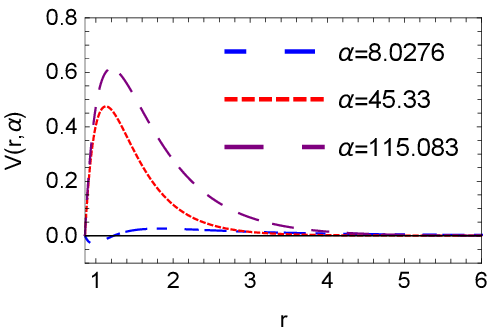}
    \hfill%
    \includegraphics{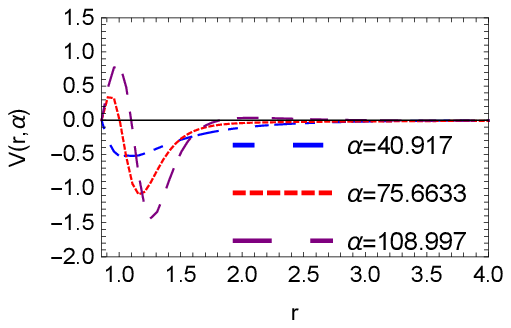}
    \hfill%
    \includegraphics{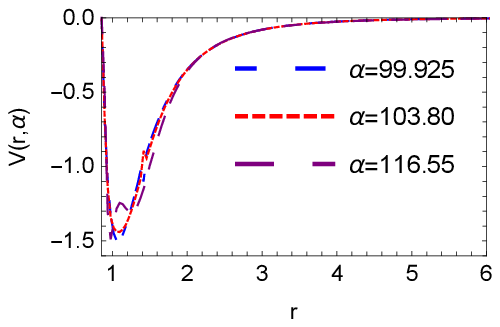}
\caption{Scalar potentials $V(r,\alpha)$  around $n=0$ (Left: $\alpha\ge8.019$), 1 (Middle: $\alpha\ge40.84$), 2 (Right: $\alpha\ge99.89$) black holes in the infinite branches for exponential coupling.  }
\end{figure*}
Actually, to determine the (in)stability of scalarized charged  black holes, we have to solve the exponential version of (\ref{perteq}) numerically by imposing an appropriate boundary condition  that a redefined scalar $\tilde{Z}(r)$ has an outgoing wave at infinity and an ingoing wave on the horizon.

Finally,  we find from Fig. 11  that the $n=0$ black hole is stable against the $l=0$-scalar because its $\Omega$ is negative, while    the $n=1, 2$ black holes  are  unstable against the $l=0$-scalar  because their $\Omega$ are positive. This indicates that there is no change on stability of scalarized charged black holes when introducing the exponential coupling to the EMCS theory.

\begin{figure*}[t!]
   \centering
   \includegraphics{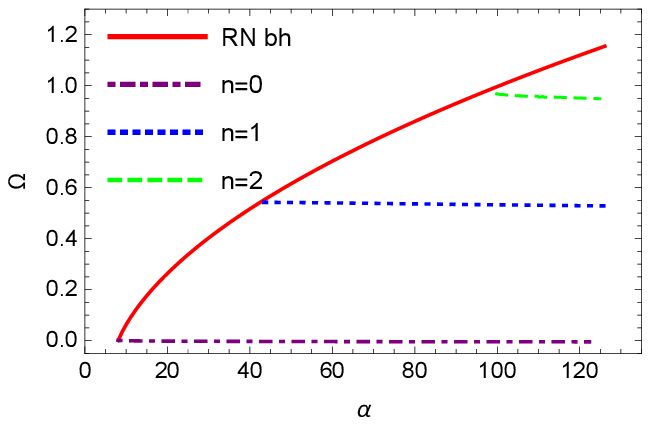}
    \hfill%
    \includegraphics{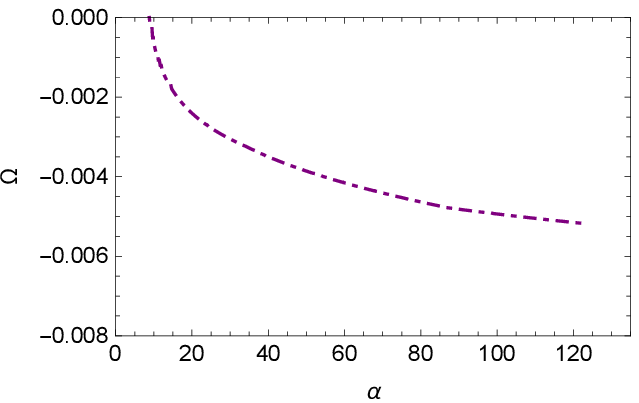}
\caption{(Left) The positive $\Omega$ as functions of  $\alpha$ for $l=0$-scalar  around the $n=1(\alpha\ge40.84),~2(\alpha\ge99.89)$ black holes found in the EMCS theory with exponential coupling. A red curve with $q=0.7$ denotes the positive $\Omega$
of $l=0$-scalar  as function of $\alpha$ around the RN black hole, indicating the unstable RN black holes for $\alpha>8.019$. (Right) The enlarged picture indicates  the small negative  $\Omega$  for $l=0$-scalar  around the $n=0(\alpha\ge8.019)$ black hole. }
\end{figure*}

\section{Discussions}
First of all, we would like to mention that  the black hole in the single branch and $n=0$ scalarized charged black hole found from the EMCS theory with  both quadratic and exponential couplings are stable against the radial  perturbations. The $n=1,2$ black holes obtained from the EMCS theory with  both quadratic and exponential couplings are unstable against the radial  perturbations.  Actually, we note that there is no difference between radial and full perturbations as far as the $s$-mode scalar perturbation is concerned.  It is worth noting that the stability results of $n=0,1,2$ black holes are consistent  with those for the EMS theory~\cite{Myung:2019oua,Myung:2018jvi}.

We summarize the stability issues for scalarized black holes obtained from three theories in Table 1.
 It implies that the inclusion of a conformally coupled  scalar does not make any changes on the stability of scalarized charged black holes except that the EMCS theory admits a stable single branch additionally.

On the other hand, the stability analysis of scalarized (charged)  black holes can replay for  whether they could be the endpoints of tachyonic instability of black holes without scalar hair. Since the $n=0$ scalarized charged black hole obtained from the EMCS theory are stable, this is  considered as an endpoint of  the RN black hole.

However, it is unclear that the scalarized charged black hole in the single branch  will  radiate to
yield the constant scalar hairy black hole  because two are  stable.

\begin{table}[h]
\begin{tabular}{|c|c|c|c|c|c|c|}
  \hline
  Theory & $n=0$  & $n=1$  & $n=2$  & Single Branch & Perturbations & Reference\\ \hline
  EGBS with EC&NA&NA&NA& U(SBH) &Radial & \cite{Doneva:2017bvd,Silva:2017uqg,Myung:2018iyq} \\ \hline
  EGBS with EC&S&U&U& NA &Radial & \cite{Blazquez-Salcedo:2018jnn} \\ \hline
   EGBS with QC&U&U&U& NA &Radial & \cite{Blazquez-Salcedo:2018jnn} \\ \hline
    EGBS with qC&S&U&U& NA &Radial & \cite{Minamitsuji:2018xde},\cite{Silva:2018qhn} \\ \hline
     EGBS with EC&S&$\cdot$&$\cdot$ & NA & Axial & \cite{Blazquez-Salcedo:2020rhf} \\ \hline
     EMS with EC&NA&NA&NA& U(RNBH) &Radial & \cite{Myung:2018vug} \\ \hline
      EMS with EC&S&U&U& NA &Full & \cite{Myung:2018jvi} \\ \hline
       EMS with QC&S&U&U& NA &Full & \cite{Myung:2019oua} \\ \hline
  EMCS with QC&S&U&U& S &Radial & This work \\ \hline
  EMCS with EC&S &U & U& S  &Radial & This work \\ \hline
\end{tabular}
\caption{Summary of stability analysis for scalarized black holes obtained from three theories.
EGBS stands for Einstein-Gauss-Bonnet-Scalar and EMS denotes Einstein-Maxwell-Scalar.
EC, QC, and  qC represent exponential coupling, quadratic coupling, quartic coupling. U(S) indicates unstable (stable) black hole.
Full implies axial+polar. SBH (RNBH) mean Schwarzschild (Reissner-Norstr\"om) black holes without scalar hair, known as the GR  solutions.
Finally, NA represents not available and `$\cdot$' implies not computed.  }
\end{table}

 \vspace{1cm}

{\bf Acknowledgments}
 \vspace{1cm}

This work was supported by the National Research Foundation of Korea (NRF) grant funded by the Korea government (MOE)
 (No. NRF-2017R1A2B4002057).

\newpage

\section*{Appendix}

\begin{appendix}

In  Appendix, we wish to derive a decoupled linearized scalar equation (\ref{nrad-eq})[=(\ref{rad-eq})].
Expanding equation (\ref{nequa1}) up to
the first order $\epsilon(\epsilon\ll1)$, we obtain the seven-coupled linearized  equations.
Four equations of  $(t,t)$, $(t, r)$, $(r,r)$, and  $(\theta,\theta)$-component are given by
\begin{eqnarray}
&&(t,r): \quad A_0 H_1+A_1\Phi+A_2\Phi'=0,\label{pertEOM1}\\
&&(t,t): \quad - \frac{A_2}{g^2}\Phi''+A_3\Phi'+A_4\Phi+A_5 H_1
-\frac{A_0}{g^2} H_1'+A_6 H_0+A_7\delta v=0,\label{pertEOM2}\\
&&(r,r): \quad -A_7 g^2\delta v-A_2g^2 \ddot{\Phi}
+A_8\Phi'+A_9\Phi+A_{10}H_1+A_0 H_0'+A_{11} H_0=0,\label{pertEOM3}\\
&&(\theta,\theta): \quad B_{0}\delta v+B_{1}H_0+B_{2}\Phi+B_{3}H_0'
+B_{4}H_1'+B_{5}\Phi' +B_{6}\Phi''\nonumber\\
&&\qquad \qquad+B_{7}H_0''+B_{8}\ddot{H}_1+B_{9}\ddot{\Phi}=0\label{pertEOM4}
\end{eqnarray}
with $g^2(r)=\frac{e^{2\delta(r)}}{N^2(r)}$.
Here, the overdot ($\dot{{}}$) denotes derivative with respect to time $t$, and $A_i(i=0,1,\cdots,11)$
and $B_i(i=0,1,\cdots,10)$ are functions of $r$ given by
\begin{eqnarray}
&&A_0=-\frac{\phi^2-3+r\phi\phi'}{3},\quad
A_1=\frac{1}{3}(\frac{2\phi}{r}+\frac{\phi N'}{N}-2\phi\delta'+4\phi'), \nonumber\\
&&A_2=-\frac{2\phi}{3},\quad A_3=\frac{e^{-2\delta}N(\phi N'-2N\phi')}{3r},\nonumber\\
&&A_4=\frac{e^{-2\delta}N}{3r^3}\left(\phi(2N-2+r N'-6e^{2\delta}\alpha r^2 v'^2)
+r(rN'\phi'+2N(3\phi'+r\phi''))\right),\nonumber\\
&&A_5=\frac{e^{-2\delta}N}{3r^2}\left(r(-3e^{2\delta}r(1+\alpha\phi^2)v'^2+N'(\phi^2-3+r\phi\phi'))\right.\nonumber\\
&&\qquad \left.+N(-3+\phi^2-r^2\phi'^2+2r\phi(2\phi'+r\phi''))\right),\nonumber\\
&&A_6=\frac{e^{-2\delta}N}{3r^2}\left(3-3rN'+\phi^2(rN'-1)+r^2\phi N'\phi'
+N(\phi^2-3-r^2\phi'^2+2r\phi(2\phi'+r\phi'')\right),\nonumber\\
&&A_7=-2N(1+\alpha\phi^2)v', \nonumber\\
&&A_8=\frac{1}{3r^2N}\left(-r\phi N'+2N(\phi(-2+r\delta')-3r\phi'\right),\nonumber\\
&&A_9=\frac{1}{3r^2N}\left(\phi(2-2rN'+6\alpha e^{2\delta}r^2v'^2+N(4r\delta'-2))
+r(-rN'+2N(r\delta'-2))\phi'\right),\nonumber\\
&&A_{10}=-\frac{\phi^2-3}{3r^2N},\quad A_{11}=-\frac{e^{2\delta}(1+\alpha\phi^2)v'^2}{N}, \nonumber\\
&&B_0=-2e^{2\delta}r^2(1+\alpha\phi^2)v',\quad B_1=e^{2\delta}r^2(1+\alpha\phi^2)v'^2,\nonumber\\
&&B_2=\frac{2\phi\left(e^{2\delta}r(9+18\alpha+\alpha \phi^4)v'^2-3N\phi'(2\phi+r\phi')\right)}{9(\phi^2-3)}
-\frac{2\phi}{3r}(N-rN'+rN\delta'),\nonumber\\
&&B_{3}=\frac{r}{12}\left(-3r(-3+\phi^2)N'+2N(3-6r\delta'+\phi^2(2r\delta'-1)-2r\phi\phi')\right),\nonumber\\
&&B_{4}=\frac{r}{12}\left(-r(-3+\phi^2)N'+2N(3-3r\delta'+\phi^2(r\delta'-1)-2r\phi\phi')\right),\nonumber\\
&&B_{5}=\frac{2r}{3}\left(-r\phi N'+N(\phi(r\delta'+1)+r\phi')\right),\quad B_{6}=-\frac{2r N\phi}{3},\nonumber\\
&&B_{7}=-\frac{r^2(\phi^2-3)N}{6},\quad B_{8}=-\frac{r^2(\phi^2-3)e^{2\delta}}{6N},\quad
B_{9}=\frac{2e^{2\delta}r\phi}{3N}.
\end{eqnarray}

Two linearized  Maxwell equations take the forms
\begin{eqnarray}
&&\nu=t: \quad A_{12}\delta v+A_{13}\delta v'+A_{14}\Phi'+A_{15}\Phi
+A_{16}(H_0'-H_1')=0,\label{pertMax1}\\
&&\nu=r:\quad A_{13}\delta\dot{v}+A_{14}\dot{\Phi}+A_{16}(\dot{H_0}-\dot{H_1})=0,\label{pertMax2}
\end{eqnarray}
where
\begin{eqnarray}
&&A_{12}=e^{2\delta}\left((1+\alpha\phi^2)(\frac{2}{r}+\delta')+2\alpha\phi\phi'\right),\quad
A_{13}=e^{2\delta}(1+\alpha\phi^2),\quad A_{14}=\frac{2\alpha e^{2\delta}\phi v'}{r},\nonumber\\
&&A_{15}=-\frac{2\alpha e^{2\delta}v'}{r^2}-\frac{2\alpha e^{2\delta}(-1+\alpha\phi^2)\phi' v'}{r(1+\alpha\phi^2)},\quad
A_{16}=-\frac{1}{2}e^{2\delta}(1+\alpha\phi^2)v'.
\end{eqnarray}
Finally, one  has a linearized  scalar  equation
\begin{eqnarray}
&&B_{10}\delta v+B_{11}H_0+B_{12}\Phi+B_{13}H_1+B_{14}H_0'+B_{15}H_1'+B_{16}\Phi'
+B_{17}\Phi''\nonumber\\
&&+B_{18}H_0''+B_{19}\ddot{H}_1+B_{20}\ddot{\Phi}=0\label{pertscal}
\end{eqnarray}
with
\begin{eqnarray}
&&B_{10}=2\alpha e^{2\delta}\phi v',\quad B_{11}=-\alpha e^{2\delta}\phi v'^2,\nonumber\\
&&B_{12}=-\frac{3N'+e^{2\delta}r\alpha(\phi^2-3)v'^2-3N\delta'}{3r^2},\quad
B_{13}=\frac{\phi}{3r^2},\nonumber\\
&&B_{14}=\frac{N\phi'}{3r}+\frac{\phi}{12}\left(3N'-4N\delta'\right)+\frac{N\phi'}{2},\nonumber\\
&&B_{15}=\frac{N\phi'}{3r}+\frac{\phi}{12}\left(N'-2N\delta'\right)+\frac{N\phi'}{2},\nonumber\\
&&B_{16}=\frac{N'-N\delta'}{r},\quad B_{17}=\frac{N}{r},\quad B_{18}=\frac{N\phi}{6},\nonumber\\
&&B_{19}=\frac{e^{2\delta}\phi}{6N},\quad B_{20}=\frac{2e^{2\delta}r\phi}{3N}.
\end{eqnarray}

Now, our main task is to diagonalize the linerized scalar equation (\ref{pertscal})  by exploiting the remaining six equations.
From (\ref{pertEOM1}), we have
\begin{eqnarray}
H_1=-\frac{A_1}{A_0}\Phi-\frac{A_2}{A_0}\Phi'\label{H1solu}
\end{eqnarray}
which implies  that $H_1$ is a redundant  field.
Integrating  (\ref{pertMax2})
with respect to $t$ leads to  $\delta v$
\begin{eqnarray}
\delta v=-\frac{A_{14}}{A_{13}}\Phi-\frac{A_{16}}{A_{13}}(H_0-H_1)\label{soludv}
\end{eqnarray}
which means that $\delta v$ is also a  redundant field.
Using (\ref{pertEOM4}), we transform   (\ref{pertscal}) to a reduced scalar equation without $H_0''$ and $\ddot{H}_1$
\begin{eqnarray}
A_{17}(H_0'+H_1')+A_{18}\Phi''-A_{18}g^2\ddot{\Phi}+A_{19}\delta v
+A_{20}\Phi'+A_{21}\Phi+A_{22}H_0+A_{23}H_1=0,\label{pertscalnew}
\end{eqnarray}
where
\begin{eqnarray}
&&A_{17}=\frac{rN^2}{36}\left((\phi^2-3)\phi+(\phi^2-9)r\phi'\right),\quad
A_{18}=\frac{r^2N^2}{18}(\phi^2-9),\quad A_{19}=-e^{2\delta}r^2(\frac{1}{3}+\alpha)N\phi v',\nonumber\\
&&A_{20}=\frac{N}{18}\left(r(\phi^2-9)N'+N(\phi^2(2-r\delta')+9r\delta'+2r\phi\phi')\right),\nonumber\\
&&A_{21}=-\frac{N}{18r(\phi^2-3)}\left(-(\phi^2-3)^2-18rN'+3e^{2\delta}r^2(1+3\alpha)(3+\phi^2)v'^2+6r\phi N'(\phi+2r\phi')\right)\nonumber\\
&&\qquad +\frac{N^2}{18r(\phi^2-3)}\left(\phi^4-2r\phi^3\phi'-6r\phi(2r\phi'-3)\phi'+r\phi^2(r\phi'^2+6\delta')
-3(r^2\phi'^2-6r\delta'-3)\right)\nonumber\\
&&\qquad+\frac{2N^2r\phi\phi''}{3(\phi^2-3)},\nonumber\\
&&A_{22}=e^{2\delta}r^2(\frac{1}{6}+\frac{\alpha}{2})N\phi v'^2,\nonumber\\
&&A_{23}=\frac{rN N'}{18}\left((\phi^2-3)\phi+(\phi^2-9)r\phi'\right)
-e^{2\delta}r^2(\frac{1}{6}+\frac{\alpha}{2})N\phi v'^2+\frac{r^2N^2}{18}(\phi^2-9)\phi''\nonumber\\
&&\qquad +\frac{N^2}{18}\left(\phi^3(1-r\delta')+r\phi^2\phi'(4-r\delta')
+9r\phi'(r\delta'-2)+\phi(3r\delta'+r^2\phi'^2-3)\right).
\end{eqnarray}

Making use of (\ref{soludv}), (\ref{H1solu}), $\delta v'$,  and  $H_1'$,
we could express (\ref{pertscalnew}) in terms of $H_0'$, $\Phi$, $\Phi'$ and $\Phi''$: called the reduced (\ref{pertscalnew}).
Combining  (\ref{pertEOM3}) with the reduced (\ref{pertscalnew}) to eliminate $H_0'$  arrives at
the master scalar equation for testing  the stability of scalarized charged black holes as
\begin{eqnarray}
\Big[g^2(r)\frac{\partial^2\Phi}{\partial t^2}\Big]-\frac{\partial^2\Phi}{\partial r^2}
+C_1(r)\frac{\partial\Phi}{\partial r}+U(r)\Phi=0, \label{rad-eq}
\end{eqnarray}
where $C_1(r)$ is expressed as
\begin{eqnarray}
C_1&=&\frac{C_2}{C_3}+\frac{C_4}{3NC_3},\nonumber\\
C_2&=&r\phi(\phi^2-3)^2(63-\phi^2+r(2\phi^2-45)\delta')\phi'
+r^2(\phi^2-3)\left(\phi^2(150-\phi^2+r(\phi^2-36)\delta')-135\right)\phi'^2\nonumber\\
&&+r^3\phi(\phi^4+84\phi^2-351)\phi'^3+r^4\phi^2(\phi^2-9)\phi'^4
+(\phi^2-3)^3(9+r(\phi^2-9)\delta'),\nonumber\\
C_3&=&9r(\phi^2-3)(\phi^2-3+r\phi\phi'),\nonumber\\
C_4&=&729(e^{2\delta}r^2v'^2-1)(r\phi'\phi-1)+81\phi^2\left(r^2(-\phi'^2
+e^{2\delta}v'^2(6-15\alpha+r^2\phi'^2))-9\right)\nonumber\\
&&+27\phi^3\phi'r\left(17+2e^{2\delta}r^2v'^2(15\alpha-4)\right)
+9\phi^4\left(27+r^2(4\phi'^2+e^{2\delta}v'^2(105\alpha-9+r^2(12\alpha-1)\phi'^2))\right)\nonumber\\
&&-9\phi^5\phi'r\left(7+e^{2\delta}v'^2r^2(28\alpha+1))
-3\phi^6(9+r^2(\phi'^2+\alpha e^{2\delta}v'^2(72+7r^2\phi'^2))\right)\nonumber\\
&&-3\phi^7\phi'r(1+4\alpha e^{2\delta}v'^2r^2)+\alpha\phi^8e^{2\delta}v'^2 r^2(9+r^2\phi'^2)
+2\alpha\phi^9 e^{2\delta}v'^2r^3\phi'+\alpha\phi^{10}e^{2\delta}v'^2r^2.
\end{eqnarray}
On the other hand, $U(r)$ takes a complicated form as
 \begin{eqnarray}
U&=&\frac{U_2}{U_1}, \nonumber\\
U_1&=&54r^2N^2\phi(\phi^2-3)^2(1+\alpha\phi^2)(-3+\phi^2+r\phi\phi'),\nonumber\\
U_2&=&D_0+\phi D_1+\phi^2D_2+\phi^3D_3+\phi^4D_4+\phi^5D_5+\phi^6D_6+\phi^7D_7+\phi^8D_8+\phi^9D_9+\phi^{10}D_{10}\nonumber\\
&&+\phi^{11}D_{11}+\phi^{12}D_{12}+\phi^{13}D_{13}
\end{eqnarray}
with
\begin{eqnarray}
D_0&=&1458r^2N^2\phi'(3\delta'+2r\phi'^2),\nonumber\\
D_1&=&-162N\left(3(1-e^{2\delta}r^2(1+3\alpha)v'^2+rN'(2-7r^2\phi'^2))\right.\nonumber\\
&&\left.+N(-3+r(-15\delta'+r(33r\delta'-35)\phi'^2+6r^3\phi'^4))\right),\nonumber\\
D_2&=&81r\phi'\left(6r^2N'^2+2N^2(8-18\alpha+r(\delta'(27\alpha-53+3r\delta')+2r(6(\alpha-2)+r\delta')\phi'^2))\right.\nonumber\\
&&\left.+N(-8+36\alpha+r(4e^{2\delta}r (6\alpha-1)v'^2-N'(-38+36\alpha+15r\delta'+2r^2\phi'^2)))\right),\nonumber\\
D_3&=&-9\left(-9r^3N'(6\alpha e^{2\delta}v'^2+rN'\phi'^2)+2N^2(9(2+3\alpha+r\delta'(18-15\alpha+r\delta'))\right.\nonumber\\
&&\left.-9r^2(-28+39\alpha+r\delta'(26-21\alpha+2r\delta'))\phi'^2+r^4(36\alpha-19)\phi'^4)\right.\nonumber\\
&&\left.+3N(rN'(-42+72\alpha-3r\delta'+2r^2(50-63\alpha+6r\delta')\phi'^2)+2(-6-9\alpha+r^2(18\alpha-5)\phi'^2\right.\nonumber\\
&&\left.+e^{2\delta}v'^2(3+3\alpha(8+27\alpha)+r(18\alpha\delta'-r(1+3\alpha)\phi'^2)))))\right),\nonumber\\
D_4&=&27r^3N'\phi'\left((18\alpha-11)N'+3e^{2\delta}r\alpha v'^2\right)\nonumber\\
&&+2N^2\left(72-270\alpha+3r\delta'(-86+147\alpha+3r(1-3\alpha)\delta')+2r^2(-31
+84\alpha+3(r+3r\alpha)\delta')\phi'^2\right)\nonumber\\
&&+3N\left(4(39\alpha-5)+6e^{2\delta}r^2\alpha v'^2(9-24\alpha+r\delta')+rN'(82-234\alpha\right)\nonumber\\
&&\left.+r(5(9\alpha-5)\delta'-2r(1+3\alpha)\phi'^2))\right),\nonumber\\
D_5&=&9r^3N'\left(3e^{2\delta}\alpha(18\alpha-17)v'^2+r(9\alpha-4)N'\phi'^2\right)+18N^2\left(54\alpha
+9r\delta'(8-18\alpha+(r-r\alpha)\delta')\right.\nonumber\\
&&\left.+r^2(65-276\alpha+r\delta'(-49+138\alpha+2r(9\alpha-4)\delta'))\phi'^2+r^4(11\alpha-1)\phi'^4\right)\nonumber\\
&&9N\left(-108\alpha+r(6r(19\alpha-2)\phi'^2+2e^{2\delta}rv'^2(-3+3\alpha(17+99\alpha)+3r(17-18\alpha)\alpha\delta'\right.\nonumber\\
&&\left.+r^2(\alpha-1)(9\alpha-1)\phi'^2)+N'(54(5\alpha-1)+9r(\alpha-1)\delta'+2r^2(31-144\alpha+2r(4-9\alpha)\delta')\phi'^2))\right),\nonumber\\
D_6&=&9r\phi'\left(r^2N'((4-33\alpha)N'+e^{2\delta}r\alpha(9\alpha-7)v'^2)+2N^2(24-162\alpha+r(\delta'(-49+222\alpha\right.\nonumber\\
&&\left.-3(r+3r\alpha)\delta')+2r(-2+24\alpha+5r\alpha\delta')\phi'^2))+
N(-16+240\alpha-2e^{2\delta}r^2v'^2(-2+\alpha(5+147\alpha)\right.\nonumber\\
&&\left.+r\alpha(-7+9\alpha)\delta')-5rN'(-10+78\alpha+r((1-15\alpha)\delta'+2r\alpha\phi'^2)))\right),\nonumber\\
\end{eqnarray}
\begin{eqnarray}
D_7&=&-3r^2N'\left(9e^{2\delta}\alpha(17\alpha-5)v'^2+r(12\alpha-1)N'\phi'^2\right)
+6N^2\left(-6+108\alpha+3r\delta'(14-72\alpha\right.\nonumber\\
&&\left.+3r(1-3\alpha)\delta')+r^2(16-207\alpha+r\delta'(-4+63\alpha+2r(12\alpha-1)\delta'))\phi'^2+r^4\alpha\phi'^4\right)\nonumber\\
&&-3N\left(12-216\alpha+r(N'(6(63\alpha-5)+9r(3\alpha-1)\delta')+2r^2(2-81\alpha+2r(1-12\alpha)\delta')\phi'^2)\right.\nonumber\\
&&\left.+2r((48\alpha-1)\phi'^2+e^{2\delta}v'^2(-3+9\alpha(7+36\alpha)+r\alpha(9(5-17\alpha)\delta'+r(51\alpha-19)\phi'^2)))\right),\nonumber\\
D_8&=&-3r\phi'\left(r^2N'(-(1+12\alpha)N'+e^{2\delta}r\alpha(21\alpha-5)v'^2)+2N^2(8-126\alpha\right.\nonumber\\
&&\left.+r(\delta'(-7+111\alpha+3r(3\alpha-1)\delta')+4r\alpha(2+r\delta')\phi'^2))+N(-4+156\alpha\right.\nonumber\\
&&\left.-2e^{2\delta}r^2\alpha v'^2(19+21\alpha+r(21\alpha-5)\delta')+rN'(6-222\alpha+r(5(1+3\alpha)\delta'-4r\alpha\phi'^2)))\right),\nonumber\\
D_9&=&3\alpha r^3N'\left(3e^{2\delta}(15\alpha-1)v'^2+rN'\phi'^2\right)+6N^2\left(30\alpha-1
+r(\delta'(3-42\alpha+(r-9\alpha r)\delta')\right.\nonumber\\
&&\left.+2\alpha r(r\delta'-4)(2+r\delta')\phi'^2)\right)+3N\left(2-60\alpha+r(-N'(2-78\alpha+\delta'(r-9r\alpha
+4\alpha r^3\phi'^2))\right.\nonumber\\
&&\left.+2r\alpha(3\phi'^2+e^{2\delta}v'^2(8+18\alpha+3r((1-15\alpha)\delta'+8r\alpha\phi'^2))))\right),\nonumber\\
D_{10}&=&-\alpha e^{2\delta}v'^2r^3\left((1-15\alpha)r N'+2N(-1-75\alpha+r(15\alpha-1)\delta')\right)\phi'\nonumber\\
&&-3r\alpha\phi'\left(-r^2N'^2-6N^2(-4+r\delta'(1+r\delta'))+N(-12+5rN'(2+r\delta'))\right),\nonumber\\
D_{11}&=&3\alpha N\left(6-rN'(6+r\delta')+2N(-3+r\delta'(3+r\delta'))\right)\nonumber\\
&&-e^{2\delta}\alpha r^2v'^2\left(r(1+9\alpha)N'-2N(1+27\alpha+r(1+9\alpha)\delta')+8\alpha r^2N\phi'^2\right),\nonumber\\
D_{12}&=&e^{2\delta}v'^2r^3\alpha^2\left(2N(r\delta'-7)-rN'\right),\nonumber\\
D_{13}&=&e^{2\delta}v'^2r^2\alpha^2\left(2N(r\delta'-3)-rN'\right).
\end{eqnarray}

Finally, we wish to mention  that two equations (\ref{pertEOM2}) and (\ref{pertMax1}) are redundant.
\end{appendix}


\newpage

\begin{thebibliography}{99}
\bibitem{Ruffini:1971bza}
  R.~Ruffini and J.~A.~Wheeler,
  Phys.\ Today {\bf 24}, no. 1, 30 (1971).
  doi:10.1063/1.3022513

\bibitem{Herdeiro:2015waa}
  C.~A.~R.~Herdeiro and E.~Radu,
  Int.\ J.\ Mod.\ Phys.\ D {\bf 24}, no. 09, 1542014 (2015)
  doi:10.1142/S0218271815420146
  [arXiv:1504.08209 [gr-qc]].

\bibitem{Bocharova:1970skc}
  N.~M.~Bocharova, K.~A.~Bronnikov and V.~N.~Melnikov,
  Vestn.\ Mosk.\ Univ.\ Ser.\ III Fiz.\ Astron.\ , no. 6, 706 (1970).

\bibitem{Bekenstein:1974sf}
  J.~D.~Bekenstein,
  Annals Phys.\  {\bf 82}, 535 (1974).
  doi:10.1016/0003-4916(74)90124-9

\bibitem{Astorino:2013sfa}
  M.~Astorino,
  Phys.\ Rev.\ D {\bf 88}, no. 10, 104027 (2013)
  doi:10.1103/PhysRevD.88.104027
  [arXiv:1307.4021 [gr-qc]].


\bibitem{Doneva:2017bvd}
  D.~D.~Doneva and S.~S.~Yazadjiev,
  Phys.\ Rev.\ Lett.\  {\bf 120}, no. 13, 131103 (2018)
  doi:10.1103/PhysRevLett.120.131103
  [arXiv:1711.01187 [gr-qc]].

\bibitem{Silva:2017uqg}
  H.~O.~Silva, J.~Sakstein, L.~Gualtieri, T.~P.~Sotiriou and E.~Berti,
  Phys.\ Rev.\ Lett.\  {\bf 120}, no. 13, 131104 (2018)
  doi:10.1103/PhysRevLett.120.131104
  [arXiv:1711.02080 [gr-qc]].

\bibitem{Antoniou:2017acq}
  G.~Antoniou, A.~Bakopoulos and P.~Kanti,
  Phys.\ Rev.\ Lett.\  {\bf 120}, no. 13, 131102 (2018)
  doi:10.1103/PhysRevLett.120.131102
  [arXiv:1711.03390 [hep-th]].


\bibitem{Herdeiro:2018wub}
  C.~A.~R.~Herdeiro, E.~Radu, N.~Sanchis-Gual and J.~A.~Font,
  Phys.\ Rev.\ Lett.\  {\bf 121}, no. 10, 101102 (2018)
  doi:10.1103/PhysRevLett.121.101102
  [arXiv:1806.05190 [gr-qc]].

\bibitem{Myung:2018vug}
  Y.~S.~Myung and D.~C.~Zou,
  Eur.\ Phys.\ J.\ C {\bf 79}, no. 3, 273 (2019)
  doi:10.1140/epjc/s10052-019-6792-6
  [arXiv:1808.02609 [gr-qc]].

\bibitem{Fernandes:2019rez}
P.~G.~Fernandes, C.~A.~Herdeiro, A.~M.~Pombo, E.~Radu and N.~Sanchis-Gual,
Class. Quant. Grav. \textbf{36}, no.13, 134002 (2019)
doi:10.1088/1361-6382/ab23a1
[arXiv:1902.05079 [gr-qc]].

\bibitem{Astefanesei:2019pfq}
D.~Astefanesei, C.~Herdeiro, A.~Pombo and E.~Radu,
JHEP \textbf{10}, 078 (2019)
doi:10.1007/JHEP10(2019)078
[arXiv:1905.08304 [hep-th]].

\bibitem{Blazquez-Salcedo:2020nhs}
J.~L.~Blázquez-Salcedo, C.~A.~Herdeiro, J.~Kunz, A.~M.~Pombo and E.~Radu,
Phys. Lett. B \textbf{806}, 135493 (2020)
doi:10.1016/j.physletb.2020.135493
[arXiv:2002.00963 [gr-qc]].



\bibitem{Zou:2019ays}
  D.~C.~Zou and Y.~S.~Myung,
  Phys.\ Lett.\ B {\bf 803}, 135332 (2020)
  doi:10.1016/j.physletb.2020.135332
  [arXiv:1911.08062 [gr-qc]].

\bibitem{Khodadi:2020jij}
  M.~Khodadi, A.~Allahyari, S.~Vagnozzi and D.~F.~Mota,
  arXiv:2005.05992 [gr-qc].

\bibitem{Dotti:2004sh}
  G.~Dotti and R.~J.~Gleiser,
  Class.\ Quant.\ Grav.\  {\bf 22}, L1 (2005)
  doi:10.1088/0264-9381/22/1/L01
  [gr-qc/0409005].

\bibitem{Blazquez-Salcedo:2018jnn}
  J.~L.~Blázquez-Salcedo, D.~D.~Doneva, J.~Kunz and S.~S.~Yazadjiev,
  Phys.\ Rev.\ D {\bf 98}, no. 8, 084011 (2018)
  doi:10.1103/PhysRevD.98.084011
  [arXiv:1805.05755 [gr-qc]].




\bibitem{Minamitsuji:2018xde}
  M.~Minamitsuji and T.~Ikeda,
  Phys.\ Rev.\ D {\bf 99}, no. 4, 044017 (2019)
  doi:10.1103/PhysRevD.99.044017
  [arXiv:1812.03551 [gr-qc]].

\bibitem{Silva:2018qhn}
  H.~O.~Silva, C.~F.~B.~Macedo, T.~P.~Sotiriou, L.~Gualtieri, J.~Sakstein and E.~Berti,
  Phys.\ Rev.\ D {\bf 99}, no. 6, 064011 (2019)
  doi:10.1103/PhysRevD.99.064011
  [arXiv:1812.05590 [gr-qc]].

\bibitem{Myung:2019oua}
  Y.~S.~Myung and D.~C.~Zou,
  Eur.\ Phys.\ J.\ C {\bf 79}, no. 8, 641 (2019)
  doi:10.1140/epjc/s10052-019-7176-7
  [arXiv:1904.09864 [gr-qc]].

\bibitem{Myung:2018jvi}
  Y.~S.~Myung and D.~C.~Zou,
  Phys.\ Lett.\ B {\bf 790}, 400 (2019)
  doi:10.1016/j.physletb.2019.01.046
  [arXiv:1812.03604 [gr-qc]].



\bibitem{Myung:2018iyq}
  Y.~S.~Myung and D.~C.~Zou,
  Phys.\ Rev.\ D {\bf 98}, no. 2, 024030 (2018)
  doi:10.1103/PhysRevD.98.024030
  [arXiv:1805.05023 [gr-qc]].



\bibitem{Blazquez-Salcedo:2020rhf}
  J.~L.~Blázquez-Salcedo, D.~D.~Doneva, S.~Kahlen, J.~Kunz, P.~Nedkova and S.~S.~Yazadjiev,
  Phys.\ Rev.\ D {\bf 101}, no. 10, 104006 (2020)
  doi:10.1103/PhysRevD.101.104006
  [arXiv:2003.02862 [gr-qc]].


\end{thebibliography}
\end{document}